\documentstyle[aps,epsf,twocolumn]{revtex}

\def\a{\alpha} \def\b{\beta} \def\g{\gamma} 
\def\d{\delta}  \def\ee{\epsilon} %\def\ee{\varepsilon}
 \def\th{\theta}  
 \def\l{\lambda}  \def\m{\mu} \def\n{\nu}
  \def\p{\pi}  \def\r{\rho}
\def\s{\sigma}   
 \def\f{\phi}   

\def\pa{\partial}   \def\half{{1\over
2}} 
\def\oa{${\cal O}(\a ')$}\def\oaa{${\cal O}(\a '^2)$}
\def\beq{\begin{equation}}\def\eeq{\end{equation}}
\def\bea{\begin{eqnarray}}\def\eea{\end{eqnarray}}
\font\cmss=cmss10 \font\cmsss=cmss10 at 7pt
\def\IZ{\relax\ifmmode\mathchoice
{\hbox{\cmss Z\kern-.4em Z}}{\hbox{\cmss Z\kern-.4em Z}}
{\lower.9pt\hbox{\cmsss Z\kern-.4em Z}}
{\lower1.2pt\hbox{\cmsss Z\kern-.4em Z}}\else{\cmss Z\kern-.4em Z}\fi}

\begin{document}
\draft
\title
      { Two-Loop Beta Functions Without Feynman Diagrams
         \thanks{This work is supported in part by
funds provided by
the U.S. Department of Energy (D.O.E.) under cooperative
research agreement \#DF-FC02-94ER40818, and by NSF Grant
PHY-92-06867. E-mail:
{\tt haagense, olsen @ctp.mit.edu, ricardos@mit.edu}.}}
\author
      {Peter E. Haagensen$^a$, Kasper Olsen$^{a,b}$, and Ricardo Schiappa$^a$}
\address
      {$^a$Center for Theoretical Physics, Laboratory for Nuclear Science and
Department of Physics, \\
Massachusetts Institute of Technology,
77 Massachusetts Avenue, Cambridge, MA 02139, USA \\
       $^b$The Niels Bohr Institute,
Blegdamsvej 17, DK-2100 Copenhagen, DENMARK}
\date {MIT-CTP\#2641, \ {\tt hep-th/9705105}}
\maketitle

\begin{abstract}
Starting from a consistency requirement between T-duality
symmetry and renormalization group flows,
the two-loop metric beta function is found for a $d\!=\!2$
bosonic sigma model on a generic, torsionless
background. The result is obtained without Feynman diagram calculations,
and represents further evidence 
that duality symmetry severely constrains renormalization flows.

\end{abstract}


\tighten

\section{Introduction}

Since the time of its discovery \cite{kikkawa}, target space duality has been
studied mostly as a symmetry of string backgrounds. That is to say, it is
realized as a transformation taking one set of fields $\{g_{\mu\nu},
b_{\mu\nu}, \phi\}$ (respectively metric, antisymmetric tensor and dilaton)
satisfying background field equations of motion, into another set
$\{\tilde{g}_{\mu\nu}, \tilde{b}_{\mu\nu}, \tilde{\phi}\}$ satisfying the
same equations of motion. As such, it represents a parameter space symmetry
of the
associated sigma model at its conformal points only. It was recently observed,
however, that it is also natural to impose it as a symmetry of the sigma
model away from conformal points, throughout the entire parameter space
\cite{haagensen}. This is expressed as the requirement
(to be made precise below)
that duality flows ``covariantly'' with the renormalization group (RG)
evolution of the background fields. Because information about the RG flow
is typically difficult to obtain, while a T-duality symmetry is considerably
easier to identify, such an interplay between duality and RG flows can be of
more than academic interest if it yields restrictions on the renormalization
patterns of the theory.

At one-loop order (${\cal O}(\alpha')$), it was shown in \cite{haagensen}
that indeed
the requirement of duality symmetry away from conformal points of the 2d
bosonic sigma model led to highly restrictive consistency conditions on the
RG beta functions of the model. It was found that these conditions
uniquely determine all beta functions at ${\cal O}(\alpha')$.
This is a particularly
striking fact, in that essentially the only condition imposed is that of
duality, a symmetry which is {\it prima facie} entirely unaware of the
renormalization structure of the model. Similar (albeit weaker) restrictions
have also been seen to follow from analogous consistency conditions in
altogether different contexts, such as the 2d Ising and Potts models
\cite{damgaard},
and the quantum Hall system \cite{burgess}.

Naturally, for sigma models, this would probably be an inconsequential
curiosity if such conditions only operated at ${\cal O}(\alpha')$. This
motivated
two of us to further investigate the consistency conditions at two-loop
order \cite{haagensenb}. For a restricted, purely metric
background, it was found that while both the beta functions
and the duality transformations are modified
by perturbative corrections, the ensuing consistency conditions (also
modified) continue nonetheless to be satisfied. This indicates that, at least
to ${\cal O}(\alpha'^2)$, duality transformations mysteriously remain informed
of the renormalization properties of the theory.

If this is so, one is led to inquire whether consistency conditions
at ${\cal O}(\alpha'^2)$ again allow for a determination of the beta functions
at that order. The purpose of the present investigation is to show that
indeed such a determination is possible.

After briefly reviewing the first nontrivial order, we will consider, as
in \cite{haagensenb}, a restricted class of backgrounds in order to probe the
consistency conditions at ${\cal O}(\alpha'^2)$. In order to be self-contained
we begin by deriving, from basic principles, the corrected duality
transformations at ${\cal O}(\alpha'^2)$ first presented in \cite{tsey}.
  From these
follow the ${\cal O}(\alpha'^2)$ consistency conditions on the beta functions
of the theory. We will then show that, out of the ten different tensor
structures possibly appearing in the two-loop beta function, only the known,
correct structure satisfies the consistency conditions. This represents
a completely independent and diagram-free determination of the two-loop beta
function of the purely metric 2d bosonic sigma model.

To be precise, with the restricted class of backgrounds we consider, this
${\cal O}(\alpha'^2)$ beta function is only determined up to a global constant.
However, it should be noted firstly that the beta function determined
is valid for entirely generic metric backgrounds and, secondly, that the
mechanism at work at ${\cal O}(\alpha')$ indicates that, had we considered a
more generic background at ${\cal O}(\alpha'^2)$, even this global constant
would have been determined.

\section{Order $\alpha'$}

We consider a $d\!=\!2$ bosonic sigma model with
a target abelian isometry ($\th\to\th$+ constant):
\begin{eqnarray}
S&=&\frac{1}{4\p\a'}\int d^2\!\s\, \left[
g_{00}(X) \pa_\a \th\pa^\a\th + 2g_{0i}(X)\pa_\a\th\pa^\a
X^i\right.\nonumber \\  && + g_{ij}(X)\pa_\a X^i\pa^\a X^j+
i\ee^{\a\b}\left. \left(
2b_{0i}(X)\pa_\a\th\pa_\b X^i\right.\right. \label{original}\\
&&\left.\left.+b_{ij}(X)\pa_\a X^i \pa_\b X^j\right)\right]\ .\nonumber
\end{eqnarray}
The adapted target space coordinates are $X^{\m}=(\th ,X^i)$,
$i=1,\ldots,D$, and
the isometry is made manifest through the independence of background tensors
on $\th$. ``Classical'' duality transformations
\cite{buscher} take a background $\{g_{\m\n},b_{\m\n}\}$ into
\begin{eqnarray}\label{duality}
\tilde{g}_{00}&=&{1\over g_{00}}\ ,\
\tilde{g}_{0i}={b_{0i}\over g_{00}}\ ,\ \tilde{b}_{0i}={g_{0i}\over
g_{00}}\ , \nonumber\\ \tilde{g}_{ij}&=&g_{ij}
-{g_{0i}g_{0j}-b_{0i}b_{0j}\over g_{00}}\ ,\\
\tilde{b}_{ij}&=&b_{ij}-{g_{0i}b_{0j}-b_{0i}g_{0j}\over g_{00}}\ .\nonumber
\end{eqnarray}
On a curved worldsheet, another background coupling must be introduced,
that of the dilaton $\f (X)$.
The RG flow of background couplings is given by their respective beta
functions:
\beq
\b^{g}_{\m\n}\equiv\m\frac{d}{d\m}g_{\m\n}\ ,\;
\b^{b}_{\m\n}\equiv\m\frac{d}{d\m}b_{\m\n}\ ,\;
\b^{\f}\equiv\m\frac{d}{d\m}\f\ ,
\eeq
while the trace of the stress energy tensor is found from the
Weyl anomaly coefficients \cite{tseytlinb}
\begin{eqnarray}
\bar{\beta}_{\mu\nu}^g&=&\beta_{\mu\nu}^g+2\alpha '\nabla_\mu
\partial_\nu\phi\ ,\nonumber\\ \bar{\beta}_{\mu\nu}^b&=&\beta_{\mu\nu}^b
+\alpha '{H_{\mu\nu}}^\lambda
\partial_\lambda\phi\label{weyl}\ ,\\ \bar{\beta}^\phi&=&\beta^\phi
+\alpha '(\partial_\mu\phi)^2\ .\nonumber
\end{eqnarray}
Both the beta functions and the Weyl anomaly coefficients will satisfy the
consistency conditions to be presented below.
However, while the latter satisfy them exactly,
the former satisfy them up to a target reparametrization
\cite{haagensen,haagensenb}. Since both
encode essentially the same RG information, for simplicity we will
consider RG motions as generated by the Weyl anomaly coefficients
in what follows.
We define (at any order) an operation $R$ on a generic functional
$F[g,b,\f]$ to be
\begin{equation}
RF[g,b,\f]={\d F\over\d g_{\m\n}}\cdot\bar{\b}_{\m\n}^g
+{\d F\over\d b_{\m\n}}\cdot\bar{\b}_{\m\n}^b
+{\d F\over\d \f}\cdot\bar{\b}^\f\ ,
\label{RF}
\end{equation}
and an operation $T$ affecting (at lowest order) the transformations
(\ref{duality})
through
\begin{equation}
TF[g,b,\f]=F[\tilde{g},\tilde{b},\tilde{\f}]
\label{TF}
\end{equation}
(where $\tilde{\f}$ will be defined shortly). 
The requirement that duality flows ``covariantly'' with the RG 
is expressed as
\begin{equation}
[T,R]=0\ .
\label{TR}
\end{equation}
When applied to (\ref{duality})
this leads to the consistency conditions first presented in
\cite{haagensen}
for the Weyl anomaly coefficients
\begin{eqnarray}
\bar{\b}^{\tilde{g}}_{00}&=&-{1\over g_{00}^2}
\bar{\b}^g_{00}\ ,\nonumber\\ \bar{\b}^{\tilde{g}}_{0i}&=&
-{1\over g_{00}^2}\left(
b_{0i}\bar{\b}^g_{00}-\bar{\b}^b_{0i}g_{00} \right)\ ,\nonumber\\
\bar{\b}^{\tilde{b}}_{0i}&=&-{1\over g_{00}^2}\left(
g_{0i}\bar{\b}^g_{00}-\bar{\b}^g_{0i}g_{00} \right)\ ,\label{consistency}\\
\bar{\b}^{\tilde{g}}_{ij}&=&\bar{\b}^g_{ij}-{1\over g_{00}}\left(
\bar{\b}^g_{0i}g_{0j}+
\bar{\b}^g_{0j}g_{0i}-\bar{\b}^b_{0i}b_{0j}
-\bar{\b}^b_{0j}b_{0i}\right)\nonumber\\
&&+ {1\over
g_{00}^2}\left( g_{0i}g_{0j}-b_{0i}b_{0j}\right) \bar{\b}^g_{00}\ ,\nonumber\\
\bar{\b}^{\tilde{b}}_{ij}&=&\bar{\b}^b_{ij}-{1\over g_{00}}\left(
\bar{\b}^g_{0i}b_{0j}+
\bar{\b}^b_{0j}g_{0i}-\bar{\b}^g_{0j}b_{0i}
-\bar{\b}^b_{0i}g_{0j}\right)\nonumber\\
&&+ {1\over
g_{00}^2}\left( g_{0i}b_{0j}-b_{0i}g_{0j}\right) \bar{\b}^g_{00}\ .\nonumber
\end{eqnarray}
At loop order $\ell$, the possible tensor structures $T_{\m\n}$ appearing
in the beta function must scale as
$T_{\m\n}(\Lambda g,\Lambda b)=\Lambda^{1-\ell}T_{\m\n}(g,b)$
under global scalings of the background fields\cite{lag}.
At \oa\ one may then have
\begin{eqnarray}
\b_{\m\n}^{g}&=&\a'\left( A\, R_{\m\n}+B\, H_{\m\l\r}H_{\n}^{\; \l\r}
+C\, g_{\m\n}R\right. \nonumber\\
&&\left. +D\, g_{\m\n}H_{\a\b\g}H^{\a\b\g}\right)\ ,\nonumber\\
\b_{\m\n}^{b}&=&\a'\left(E\, \nabla^{\l}H_{\m\n\l}\right)\ ,
\label{tensorone}
\end{eqnarray}
with $A,B,C,D,E$ being determined from one-loop Feynman diagrams. As found in
\cite{haagensen}, requiring (\ref{consistency}) to be satisfied,
and choosing $A=1$ determines
$B=-1/4$, $E=-1/2$, and $C=D=0$,
independently of any diagram calculations. As it
turns out, the consistency conditions (\ref{consistency}) on
$g_{\m\n}$ and $b_{\m\n}$ alone also
allows for an independent determination of the dilaton
transformation (or ``shift'') $\tilde{\f}=
\f-{1\over2}\ln g_{00}$. Applying (\ref{TR}) to this then yields
the dilaton beta function \cite{haagensenb}.

\section{Order $\alpha'^2$}

At the next order $R$ is modified by the two-loop beta functions, and
one must determine the appropriate modifications in $T$ such that
$[T,R]=0$ continues to hold. We work at this order with restricted
backgrounds of the form
\beq\label{metricrest}
g_{\mu\nu}=\left( \begin{array}{cc}a & 0\\ 0 &\bar{g}_{ij}
\end{array}\right)\ ,
\eeq
and $b_{\m\n}\!=\!0$, so that no torsion appears in the dual background
either. It is useful to define at this point the following two
quantities:
$a_{i}\equiv\pa_{i}\ln a$, and $q_{ij}\equiv\bar{\nabla}_ia_j+\half a_ia_j$,
where barred quantities here and below refer to the metric $\bar{g}_{ij}$
(also, indices $i,j,\ldots$, are contracted with the metric $\bar{g}_{ij}$).
Within this class of backgrounds classical duality transformations
reduce to the operation $a\to 1/a$, and it is simple to determine the
possible corrections to $T$ from a few basic requirements: {\it i)}
$\tilde{g}_{ij}=g_{ij}=\bar{g}_{ij}$ does not get modified, as it
corresponds to sigma model couplings entirely disconnected from the path
integral dualization procedure (cf. \cite{buscher}); {\it ii)} corrections
should be $D$-dimensional generally covariant; {\it iii)} corrections to
$\tilde{a}=1/a$ must be proportional to $a_i$:
\beq\label{acorrections}
\ln\tilde{a}=-\ln a+\alpha' m_{i}a^{i},\ m_i=m_i(a,\bar{g}_{ij})\ ,\eeq
as it is simple to see that
classical consistency conditions would be satisfied for $a=$ constant;
{\it iv)} dimensional analysis: $[\a ']=L^2$ and $[a_i]=1/L$, where $L$
is a target length, so that $[m_i]=1/L$; {\it v)} $m_i$ should not contain
nontrivial denominators, as the corrections should be finite for finite
geometries; {\it vi)} because the duality group should still be $\IZ_2$, by
applying the transformations (\ref{acorrections}) twice one should re-obtain
the original model. This constrains $m_i$ to be odd under classical
duality:
\beq
\tilde{m}_i\equiv m_i(1/a,\bar{g}_{ij})=-m_{i}(a,\bar{g}_{ij})\ .
\eeq
All of the above then yields
\beq
m_{i}=\l\, a_i\ ,
\eeq
with $\l$ an undetermined real constant. As discussed in \cite{haagensenb},
moreover, we shall also require the measure factor $\sqrt{g}\exp{(-2\f)}$
to be invariant (so that $[T,R]=0$ implies invariance of the string
background effective action), thus fixing also the correction on the
dilaton transformation to be 1/4 that of $g_{00}$. Altogether, for
the backgrounds (\ref{metricrest}) the corrected duality transformations
are:
\bea
\ln\tilde{a}&=&-\ln a+\l\a' a_{i}a^{i}\ ,\nonumber\\
\tilde{g}_{ij}&=&g_{ij}=\bar{g}_{ij}\ ,\label{tlduality}\\
\tilde{\phi}&=&\phi -\frac{1}{2}\ln a +\frac{\l}{4}\a' a_{i}a^{i}\ .
\nonumber
\eea
The consistency conditions again follow by applying $R$ to the
above and using $[T,R]=0$ on the l.h.s.:
\bea
\frac{1}{\tilde{a}}\tilde{\bar{\beta}}_{00} &=&
-\frac{1}{a}\bar{\beta}_{00}+2\l\alpha'\left[
a^{i}\partial_{i}\left( \frac{1}{a}\bar{\beta}_{00}\right)
-\frac{1}{2} a^{i}a^{j}\bar{\beta}_{ij}\right]\ ,\nonumber\\
\tilde{\bar{\beta}}_{ij}&=& \bar{\beta}_{ij}\ ,\label{tlconsistency}\\
\tilde{\bar{\beta}}^{\phi}&=& \bar{\beta}^\phi-\frac{1}{2a}
\bar{\beta}_{00}+\frac{\l}{2}\alpha'\left[ a^{i}\partial_{i}\left(
\frac{1}{a}\bar{\beta}_{00}\right)
-\frac{1}{2} a^{i}a^{j}\bar{\beta}_{ij}\right]\ .\nonumber
\eea
The terms scaling correctly under $g\rightarrow \Lambda g$ at this order,
and thus possibly present in the beta function, are
\bea
\b_{\m\n}^{(2)} &=& A_1\, \nabla_\mu\nabla_\nu R
+A_2\,\nabla^2 R_{\mu\nu}+
A_3\, R_{\mu\alpha\nu\beta}R^{\alpha\beta} \nonumber\\
&&+A_4\, R_{\mu\alpha\beta\gamma}{R_\nu}^{\alpha\beta\gamma}+
A_5\, R_{\mu\alpha}{R_\nu}^\alpha+A_6\, R_{\mu\nu}R\label{tltensors}\\
&&+A_7\, g_{\mu\nu}\nabla^2 R +A_8\, g_{\mu\nu}R^2 +
A_9\, g_{\mu\nu}R_{\alpha\beta}R^{\alpha\beta}\nonumber\\
&&+A_{10}\,g_{\mu\nu}
R_{\alpha\beta\gamma\delta}R^{\alpha\beta\gamma\delta}
\nonumber
\eea
(we have used Bianchi identities to reduce from a larger set of tensor
structures). It will suffice in fact to study the consistency conditions
for the $(ij)$ components, $\tilde{\bar{\b}}_{ij}=\bar{\b}_{ij}$, in order
to determine the only structure satisfying all the consistency
conditions.

We write
\beq
\bar{\b}_{ij}=\a'\left(\b_{ij}^{(1)}+2\bar{\nabla}_{i}\pa_{j}\f\right)
+\a'^2\b_{ij}^{(2)}\ ,
\eeq
where $\b_{ij}^{(1)}\!=\!R_{ij}\!=\!\bar{R}_{ij}-\frac{1}{2}q_{ij}$
is the one-loop beta
function, and perform the duality transformation (\ref{tlduality}),
keeping terms to order ${\cal O}(\a'^2)$. Using the fact that the
one-loop Weyl anomaly coefficient satisfies the one-loop consistency
conditions (\ref{consistency}), we arrive at
\beq
\tilde{\b}_{ij}^{(2)}=\b_{ij}^{(2)}-\frac{1}{4}\l a_{(i}\pa_{j)}
(a^ka_k)\ ,
\eeq
where the duality transformation of $\b_{ij}^{(2)}$ is given simply
by $a \rightarrow 1/a$ without $\a'$ corrections, since this is already
${\cal O}(\a'^2)$. Separating the possible tensor
structures into even and odd tensors under $a \rightarrow 1/a$,
\beq
\b_{ij}^{(2)}=E_{ij}+O_{ij}\ , \; \; \;
\tilde{E}_{ij}=E_{ij}\ , \; \; \; \tilde{O}_{ij}=-O_{ij}\ ,
\eeq
the even structures drop out and we are left with
\beq\label{odd}
O_{ij}=\frac{1}{8}\l a_{(i}\pa_{j)}(a^ka_k)\ .
\eeq
We now perform a standard Kaluza-Klein reduction on the ten terms in
(\ref{tltensors}) to identify which if any satisfy this condition.
The results can be obtained using the formulas in the Appendix of
\cite{haagensenb}, and are as follows:
\bea
&(1)&:\ \nabla_{i}\nabla_{j}R = \bar{\nabla}_{i}\bar{\nabla}_{j}
(\bar{R}-{q_ n}^{n})\ ,\nonumber\\
&(2)&:\ \nabla^{2}R_{ij} = (\bar{\nabla}^{2}
+\frac{1}{2}a_{k}\bar{\nabla}^{k})
(\bar{R}_{ij}-\frac{1}{2}q_{ij})
-\frac{1}{4}a_{i}a_{j}{q_ n}^{n}\nonumber\\
&&\hspace{18mm}-\frac{1}{4}a^{k}a_{(i}\left(\bar{R}_{j)k}-\frac{1}{2}q_{j)k}
\right)\ ,\nonumber\\
&(3)&:\ R_{i\alpha j\beta}R^{\alpha\beta} = \frac{1}{4}q_{ij}{q_ n}^{n}
+\bar{R}_{injm}(\bar{R}^{nm}-\frac{1}{2}q^{nm})\ ,\nonumber\\
&(4)&:\ R_{i\alpha\beta\gamma}{R_ j}^{\alpha\beta\gamma} =
\frac{1}{2}q_{ik}{q_ j}^{k}+\bar{R}_{iknm}\bar{R_ j}^{knm}\ ,\nonumber\\
&(5)&:\ R_{i\alpha}{R_j}^\alpha =
\bar{R}_{ik}{\bar{R}_j}^{\;\, k}-{1 \over 2}\bar{R}_{k(i}{q_{j)}}^k+
{1 \over 4}q_{ik}{q_j}^k\ ,\label{kaluza}\\
&(6)&:\ R_{ij}R = (\bar{R}_{ij}-{1 \over 2}q_{ij})(\bar{R}-{q_n}^n)\ ,
\nonumber\\
&(7)&:\ g_{ij}\nabla^{2}R = \bar{g}_{ij}\left[ \half a^{k}\partial_{k}
(\bar{R}-{q_ m}^{m})\right.\nonumber\\
&&\hspace{21mm}\left. \phantom{ {a\over b}}
+\bar{\nabla}^{k}\partial_{k}(\bar{R}-{q_ m}^{m})
\right]\ ,\nonumber\\
&(8)&:\ g_{ij}R^{2} = \bar{g}_{ij}\left(\bar{R}-{q_ m}^{m}\right)^{2}\ ,
\nonumber\\
&(9)&:\ g_{ij}R_{\alpha\beta}R^{\alpha\beta} = \bar{g}_{ij}
\left[ \frac{1}{4}({q_ m}^{m})^{2}+(\bar{R}_{km}-\frac{1}{2}q_{km})^{2}
\right]\ ,\nonumber\\
&(10)&:\ g_{ij}R_{\alpha\beta\gamma\delta}R^{\alpha\beta\gamma\delta}
 =  \bar{g}_{ij}\left[ q_{km}q^{km}+\bar{R}_{k\ell mn}
\bar{R}^{k\ell mn}\right]\ .\nonumber
\eea
The respective odd parts are
\bea
O_{ij}^{(1)}&=&-\bar{\nabla}_{i}\bar{\nabla}_{j}
\bar{\nabla}_{n}a^{n}\ ,\nonumber\\
O_{ij}^{(2)}&=&\frac{1}{2}a_{k}\bar{\nabla}^{k}\bar{R}_{ij}
-\frac{1}{2}\bar{\nabla}^{2}\bar{\nabla}_{i}a_{j}
-\frac{1}{4}a_{i}a_{j}\bar{\nabla}_{k}a^{k}\ ,\nonumber\\
O_{ij}^{(3)}&=&-\frac{1}{2}\bar{R}_{injm}\bar{\nabla}^{n}a^{m}
+\frac{1}{8}a_{n}a^{n}\bar{\nabla}_{i}a_{j}
+\frac{1}{8}a_{i}a_{j}\bar{\nabla}_{n}a^{n}\ ,\nonumber\\
O_{ij}^{(4)}&=&\frac{1}{4}a_{k}a_{(i}\bar{\nabla}_{j)}a^{k}\ ,\nonumber\\
O_{ij}^{(5)}&=&-{1 \over 2}\bar{R}_{k(i}\bar{\nabla}_{j)}a^k+
\frac{1}{8}a_{k}a_{(i}\bar{\nabla}_{j)}a^{k}\ ,\label{odd2}\\
O_{ij}^{(6)}&=&-{1\over2}\bar{R}\bar{\nabla}_ia_j-\bar{R}_{ij}
\bar{\nabla}_na^n+{1\over4}a_ia_j\bar{\nabla}_na^n\nonumber\\
&&+{1\over4}a_na^n
\bar{\nabla}_ia_j\ ,\nonumber\\
O_{ij}^{(7)}&=&\bar{g}_{ij}\left[ \frac{1}{2}a^{k}\partial_{k}
(\bar{R}-\frac{1}{2}a_{m}a^{m})-\bar{\nabla}^{k}\partial_{k}
(\bar{\nabla}_{m}a^{m})\right]\ ,\nonumber\\
O_{ij}^{(8)}&=& \bar{g}_{ij}\left[-2(\bar{\nabla}^{k}a_{k})\bar{R}
+(\bar{\nabla}^{k}a_{k})a^{m}a_{m}\right]\ ,\nonumber\\
O_{ij}^{(9)}&=&\bar{g}_{ij}\left[ \frac{1}{4}(\bar{\nabla}^{k}a_{k})a^{m}a_{m}
-(\bar{\nabla}_{k}a_{m})\bar{R}^{km}\right.\nonumber\\
&&\; \; \; \; \;\left. +\frac{1}{4}
(\bar{\nabla}_{k}a_{m})a^{k}a^{m}
\right]\ ,\nonumber\\
O_{ij}^{(10)}&=& \bar{g}_{ij}
(\bar{\nabla}_{k}a_{m})a^{k}a^{m}\ .\nonumber
\eea
It is fortunate that none of these tensors contain purely
even structures, since such structures are left unconstrained (and thus
undetermined) by duality. The only odd term of the form
(\ref{odd}) comes from $A_{4}R_{\m\alpha\beta\gamma}
{R_\n}^{\alpha\beta\gamma}$, and
a detailed inspection shows that no linear combination of the
other terms gives rise to odd tensors generically of the form (\ref{odd}). This
determines that, with the requirement of covariance of duality under
the RG, the \oaa\ term in the beta function is
\beq\label{xxx}
\beta^{(2)}_{\mu\nu}=\lambda R_{\mu\alpha\beta\gamma}{R_\nu}^
{\alpha\beta\gamma}\ .
\eeq
One should now check that the corresponding (00) component also
satisfies its consistency condition.
A straightforward computation shows that it does,
and the determination of the two-loop beta function is thus complete.

Although we treated a restricted class of metric
backgrounds, our result is valid for a generic metric, since none of
the possible tensor structures are built out of the off-block-diagonal
$g_{0i}$ elements alone (in which case our consistency conditions would
be blind to them, just as they are to the even terms $E_{ij}$).

Some final comments on scheme dependence are also in order: for a
purely metric background, it is well-known \cite{metsaev}
that the two-loop beta function
is scheme independent within the standard set of subtraction schemes
determined by minimal and nonminimal subtractions of the one-loop divergent
structure $R_{\m\n}$. Under a broader definition of subtraction
scheme, however, when other terms may also be subtracted, {\it e.g.}, 
of the type $g_{\m\n}R$, 
then the beta function becomes scheme dependent, and differs from 
(\ref{xxx}). Our duality constraints have determined a beta function 
falling into the first (and standard) class of schemes,
{\it i.e.}, those in which the one-loop subtractions are of the form
$({\rm const.}+1/\ee )R_{\m\n}$. This is natural to expect,
as these represent the subtraction of the inherent divergence of the 
theory. However, it raises the question of whether the duality 
constraints clash against the possibility of making
more general subtractions.
We have recently found \cite{haagensenc} that in fact there is 
no clash, since it is possible to explicitly determine the 
modification in the
duality transformations themselves under a field redefinition, 
and they will be such as to preserve the consistency
conditions w.r.t. the redefined beta functions.
The statement $[T,R]=0$ thus acquires a meaning beyond and
independent of any field redefinition ambiguity. 

Simply using the requirements that duality and the RG
commute as motions in the parameter space of the sigma model, we have
been able to determine the two-loop beta function to be
\beq
\beta_{\mu\nu}=\alpha' R_{\mu\nu}+{\alpha'}^2\lambda R_{\mu\alpha\beta
\gamma}{R_\nu}^{\alpha\beta\gamma}\ ,
\eeq
for an entirely generic metric background, without any Feynman diagram
calculations. Because we used an extremely restrictive class of
backgrounds, it was not possible to determine the value of $\lambda$
(the correct value is $\lambda={1\over2}$). However, we expect that,
similarly to what happens at ${\cal O}(\alpha')$, once a more generic
background is used in the consistency conditions, even this constant
should be determined.

That duality symmetry should yield information on the renormalization
structure of a theory is to us a striking fact, and one which we intend
to further explore in the future.

\acknowledgements

One of us (RS) is partially supported by the Praxis XXI grant BD-3372/94
(Portugal).

\end{document}